%% file: main.tex
\documentclass[sigconf]{acmart}
\setcopyright{none}
\renewcommand\footnotetextcopyrightpermission[1]{} % removes footnote with conference information in first column
\usepackage{multirow}
\usepackage{caption}
\usepackage{subcaption}
\usepackage{graphicx}
\usepackage{float} 
\usepackage{enumitem}
\usepackage{algorithm} 
\usepackage{algorithmic}
\usepackage{booktabs} 
\usepackage{threeparttable}
\usepackage{makecell}

\AtBeginDocument{%
  \providecommand\BibTeX{{%
    \normalfont B\kern-0.5em{\scshape i\kern-0.25em b}\kern-0.8em\TeX}}}

\begin{document}

\title{LIBER: Lifelong User Behavior Modeling Based on Large Language Models }

% \authornote{Co-first authors with equal contributions.}

\newcommand{\chenxu}[1]{{\bf \color{red} [[Chenxu says ``#1'']]}}

\author{Chenxu Zhu}
\authornote{Co-first authors with equal contributions.}
\email{zhuchenxu1@huawei.com}
\affiliation{Huawei Noah's Ark Lab\country{China}}

\author{Shigang Quan}
\authornotemark[1]
\email{quan123@sjtu.edu.cn}
\affiliation{Shanghai Jiao Tong University\country{China}}

\author{Bo Chen}
\email{chenbo116@huawei.com}
\affiliation{Huawei Noah's Ark Lab\country{China}}

\author{Jianghao Lin}
\email{chiangel@sjtu.edu.cn}
\affiliation{Shanghai Jiao Tong University\country{China}}

\author{Xiaoling Cai}
\email{caixiaoling2@huawei.com}
\affiliation{Consumer Business Group, Huawei\country{China}}

\author{Hong Zhu}
\email{zhuhong8@huawei.com}
\affiliation{Consumer Business Group, Huawei\country{China}}

\author{Xiangyang Li}
\email{lixiangyang34@huawei.com}
\affiliation{Huawei Noah's Ark Lab\country{China}}

\author{Yunjia Xi}
\email{xiyunjia@sjtu.edu.cn}
\affiliation{Shanghai Jiao Tong University\country{China}}

\author{Weinan Zhang}
\email{wnzhang@sjtu.edu.cn}
\affiliation{Shanghai Jiao Tong University\country{China}}

\author{Ruiming Tang}
\authornote{Corresponding author.}
\email{tangruiming@huawei.com}
\affiliation{Huawei Noah's Ark Lab\country{China}}

\renewcommand{\shortauthors}{Chenxu Zhu, et al.}

\input{section/abstract}

\maketitle

\input{section/introduction}
\begin{figure*}[t]
  \centering
  \vspace{-0.8em}
  \includegraphics[width=0.98\linewidth]{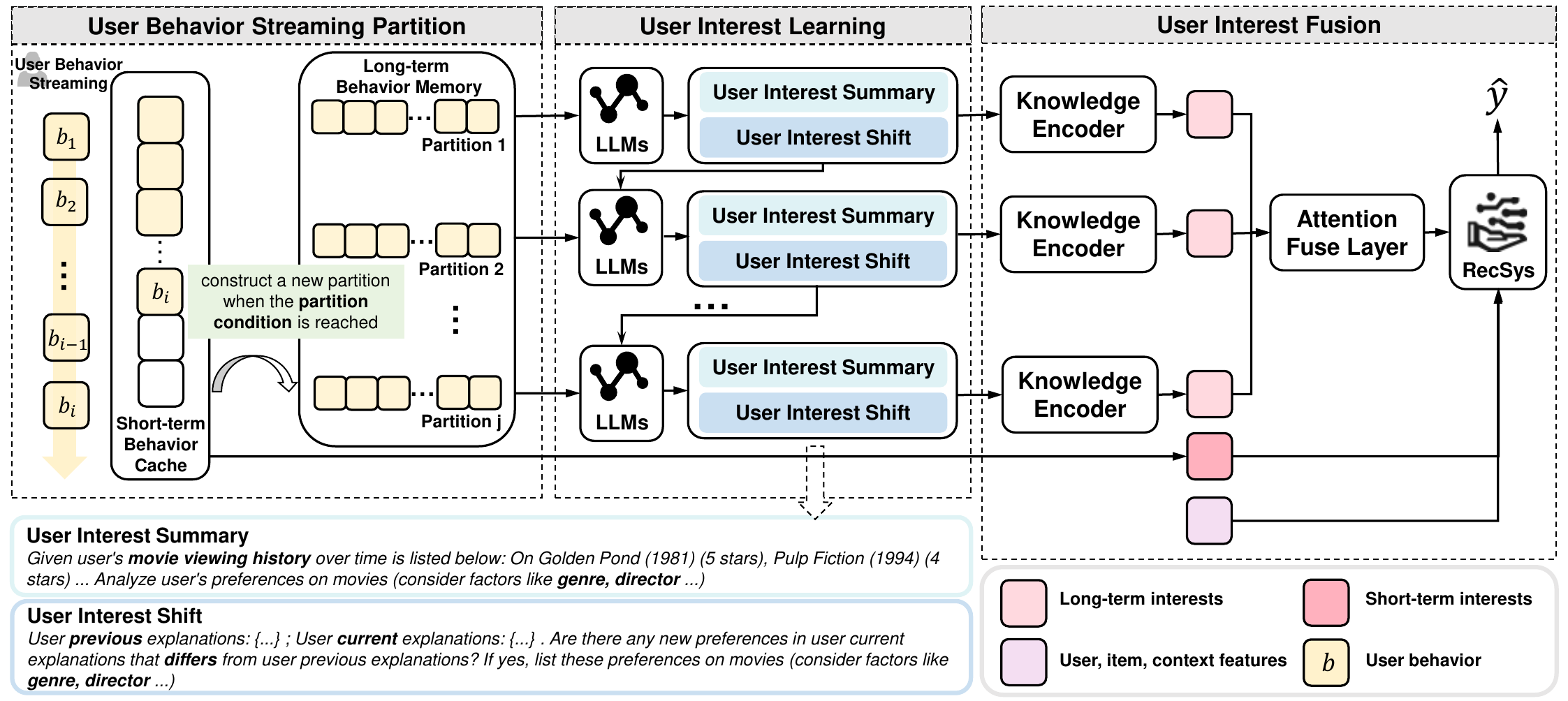}
  \vspace{-1.0em}
  \caption{The overview architecture of LIBER. LIBER consists of three modules: User Behavior Streaming Partition Module, User Interest Learning Module, and User Interest Fusion Module.}
  \vspace{-1.0em}
  \label{fig: model}
\end{figure*}

\input{section/relatedwork}

\input{section/methodology}

\input{section/experiment}

\input{section/conclusion}

\bibliographystyle{ACM-Reference-Format}
\bibliography{sample-base}

\appendix
\input{section/appendix}

\end{document}

%% file: section/abstract.tex
\begin{abstract}

Click-Through Rate (CTR) prediction plays a vital role in recommender systems. Recently, large language models (LLMs) have been applied in recommender systems due to their emergence abilities. 
While leveraging semantic information from LLMs has shown some improvements in the performance of recommender systems, two notable limitations persist in these studies.
First, LLM-enhanced recommender systems encounter challenges in extracting valuable information from lifelong user behavior sequences within textual contexts for recommendation tasks. 
Second, the inherent variability in human behaviors leads to a constant stream of new behaviors and irregularly fluctuating user interests. 
This characteristic imposes two significant challenges on existing models. On the one hand, it presents difficulties for LLMs in effectively capturing the dynamic shifts in user interests within these sequences, and on the other hand, there exists the issue of substantial computational overhead if the LLMs necessitate recurrent calls upon each update to the user sequences. In this work, we propose \textbf{Li}felong User \textbf{Be}havio\textbf{r} Modeling (LIBER) based on large language models, which includes three modules:
(1) User Behavior Streaming Partition (UBSP), (2) User Interest Learning (UIL), and (3) User Interest Fusion (UIF). 
Initially, UBSP is employed to condense lengthy user behavior sequences into shorter partitions in an incremental paradigm, facilitating more efficient processing. Subsequently, UIL leverages LLMs in a cascading way to infer insights from these partitions. Finally, UIF integrates the textual outputs generated by the aforementioned processes to construct a comprehensive representation and then this representation can be incorporated by any recommendation model to enhance the performance.
LIBER has been deployed on Huawei's music recommendation service and achieved substantial improvements in users' play count and play time by $3.01\%$ and $7.69\%$, respectively.

%We also conduct experiments on three large-scale datasets to show the superiority of our method compared with the state-of-the-art baselines.

\end{abstract}

%% file: section/introduction.tex
\section{INTRODUCTION}\label{sec:intro}
Click-Through Rate (CTR) prediction~\cite{deepfm,dcn,dcnv2,din,dien}, which aims to predict the probability of the user clicking on the recommended items (\emph{e.g.}, music, advertisement), plays a core role in recommender systems. Recently, large language models (LLMs)~\cite{glm,gpt,llama,llama2} have achieved remarkable breakthroughs and shown amazing emergence capabilities, thus giving a mushroom growth to the promising research direction of LLM-enhanced recommender systems~\cite{p5,tallrec,rella,kar,clickprompt,lin2023can,flip} for performance enhancements. 

User behavior sequence modeling is especially important for CTR prediction due to the rich user interest information contained in the behaviors~\cite{din,dien}. However, traditional methods for user behavior sequence modeling~\cite{ubr,sim,mamba4rec} only consider the information from their own datasets, ignoring other valuable external knowledge. 
To overcome this problem, researcher~\cite{kar,rella} introduces LLMs to assist user behavior sequence modeling, because LLMs have accumulated a great amount of factual knowledge from the web and possess strong reasoning capabilities to better analyze users' interests.
Specifically, KAR~\cite{kar} employs LLMs to summarize the user sequences while Rella~\cite{rella} retrievals the most relevant behaviors and directly makes the predictions by LLMs.

\begin{figure}[t]
\vspace{-0.8em}
\includegraphics[width=0.48\columnwidth]{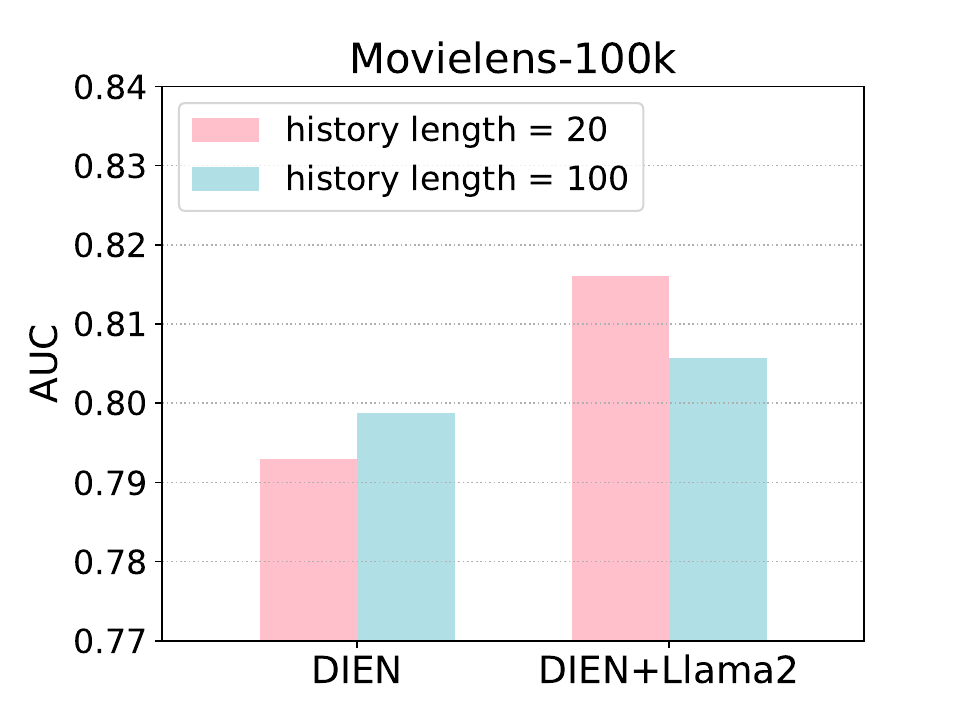} 
\includegraphics[width=0.48\columnwidth]{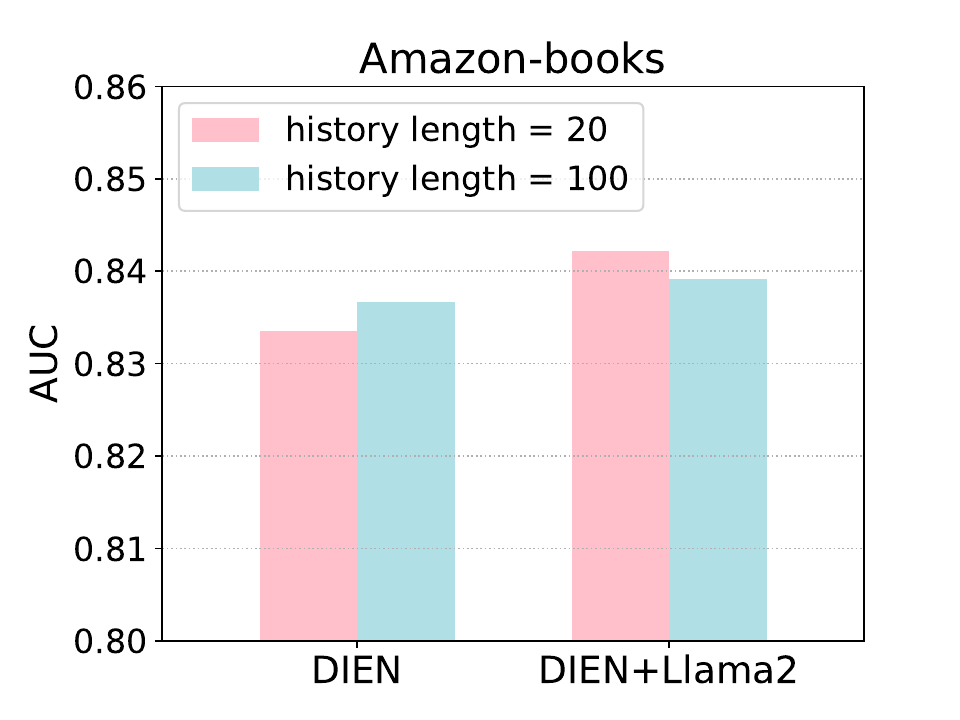}   
\vspace{-1.0em}
\caption{The illustration of the lifelong user behavior incomprehension problem for LLMs. We report the AUC performance of DIEN and DIEN+Llama2 with different history lengths on Movielens-100k and Amazon-book datasets. DIEN is a traditional recommendation model that only utilizes ID-based information, whereas DIEN+Llama2 is an LLM-enhanced model and utilizes both ID-based and textual information, which first uses Llama2 to exploit the textual information of behavior sequences and then employs the output of Llama2 as an additional feature for DIEN. It is observed that for both two datasets, as the length of behavior sequence $K$ grows, the performances of DIEN are improved but the performances of DIEN+Llama2 decrease significantly.} 
\vspace{-1.0em}
\label{fig:motivation}
\end{figure}

However, there exist two significant limitations in the existing LLM-enhanced CTR prediction works:
 \textbf{(1)} LLMs fail to extract textual context information of lifelong user behavior sequences for recommendation tasks~\cite{rella}, which we refer to as the lifelong user behavior incomprehension problem. This problem is shown in Figure~\ref{fig:motivation}, where DIEN~\cite{dien} is a traditional recommendation model while DIEN+Llama2 is a LLM-enhanced model. DIEN+Llama2 first utilizes Llama2~\cite{llama2} (a popular open-source LLM with a context window of 4096 tokens) to exploit the textual information of user behavior sequences and then employs it as an additional feature for DIEN. As we can observe, DIEN enjoys performance gains as the length of involved user behavior sequence $K$ grows. However, the performance of DIEN+Llama2 has an obvious drop with a larger $K$ value. 
 Even though the number of involved tokens (around $2000$ tokens when $K=100$) is far less than the context window limitation ($4096$ tokens), LLMs struggle in effectively capturing the highlights~\cite{lost_in_the_middle,rella} when $K$ grows to a certain extent.
 Besides, the user behavior sequence required for industrial environments is generally longer, resulting in the possibility of using tokens that exceed the context window limitation of LLMs (\textit{i.e.}, more than $20000$ tokens when $K>=1000$).
 \textbf{(2)} The existing methods do not account for the inherent variability in behaviors, which results in a constant stream of new behaviors and irregularly fluctuating user interests. This leads to two serious challenges. First, since the current LLM-enhanced works tend to treat each item in user behavior sequences equally without considering its obsolescence, it is difficult for LLMs to understand the dynamic variations of user interests. Second, LLMs need to be re-executed whenever the user behavior sequence is updated, which brings huge computational costs.

To solve the aforementioned problems, we propose a simple yet effective framework called \textbf{Li}felong User \textbf{Be}havio\textbf{r} Modeling (LIBER) based on large language models. Inspired by the computer data storage technique~\cite{cachememory}, we divide the lifelong sequence of user behaviors into short-term behavior cache and long-term behavior memory in LIBER. Long-term behavior memory is conducted in partitions, with new partitions being increased gradually only if certain conditions are met. 
LIBER performs LLM enhancement only for each partition in long-term behavior memory while using the traditional sequence modeling method to deal with short-term behavior cache. Since the past partitions for each user are fixed, each partition only needs to perform LLMs once. Therefore, the efficiency is improved greatly. Besides, LIBER restricts the user behaviors in each partition to a suitable length to deal with the lifelong user behavior incomprehension problem. Furthermore, LIBER introduces a cascading paradigm that considers the association of different partitions to learn the evolution of user interests.

Specifically, LIBER consists of three modules: (1) User Behavior Streaming Partition (UBSP), (2) User Interest Learning (UIL), and (3) User Interest Fusion (UIF). Whenever a new user behavior is received, it is first allocated to the short-term behavior cache in UBSP. When the specific partition condition is reached, these behaviors in the cache are employed to construct a new partition in the long-term behavior memory. Each partition is processed by UIL to generate an intra-block user interest summary and an inter-block user interest shift in a cascading way. The user interest summary and the user interest shift are passed to a knowledge encoder and fused by an attention layer in UIF. Finally, this fusion representation, which implies long-term user interest, can be incorporated by any recommendation model to enhance the performance. 
% By applying our proposed LIBER, we can achieve the best performance on three large-scale datasets (two public, and one industrial).

In conclusion, our contributions can be summarized as follows:

\begin{itemize}[leftmargin = 10 pt]
    \item To handle the lifelong user behavior incomprehension problem, we present a novel incremental framework called LIBER to model the lifelong user behavior sequences with large language models, simultaneously improving the effectiveness and efficiency.
    \item The proposed LIBER with three core stages (namely, User Behavior Streaming Partition, User Interest Learning, and User Interest Fusion) automatically explores the user interest summary and user interest shift in a cascading paradigm to assist the recommendation.
    % \item Offline experiments on two public datasets and one industrial dataset demonstrate the superior performance of LIBER compared with the existing models. Besides, as a model-agnostic framework, we validate the compatibility of LIBER for different backbone recommendation models and different large language models. 
    \item Offline experiments on two public datasets and one industrial dataset demonstrate the superior performance of LIBER. Besides, an online A/B test further confirms the effectiveness and applicability of LIBER. 
\end{itemize}

%% file: section/relatedwork.tex
\section{RELATED WORK}
\subsection{Traditional CTR Prediction Model}
CTR prediction task is to predict the probability of clicking the target item based on input context information~\cite{deepfm,din,dcn,guo2023dffm}. The main methods of CTR prediction can be divided into two types: feature interaction models~\cite{dcn,dcnv2,zhu2021aim,xdeepfm} and user behavior sequence models~\cite{din,dien,chen2022efficient}.
The feature interaction models aim to cross features through different operations (such as factorization machine~\cite{fm}, cross-net~\cite{dcn,dcnv2}, attention~\cite{autoint}, and so on). DeepFM~\cite{deepfm} and xDeepFM~\cite{xdeepfm} are DNN-based factorization models to learn both low-order and high-order feature interactions. DCN~\cite{dcn} and DCNV2~\cite{dcnv2} propose the cross network to enable feature interactions and use stacked layers to learn feature interactions of different orders. AutoInt~\cite{autoint} and FiBiNet~\cite{fibinet} adopt the attention-based mechanism to explicitly construct feature interactions, which provide explainable attention weights.
Another line is to extract information from user behavior sequences. The user behavior sequences are usually sequential ID features, so different structures (attention~\cite{din,sasrec}, GRU \cite{dien}, CNN~\cite{caser}, and so on) are designed to model these features. Among these methods, DIN~\cite{din} utilizes an attention mechanism to represent the relation between the target item and historical items. DIEN \cite{dien} models interest evolving through attention-based GRU. Nevertheless, all these models neglect the powerful abilities of large language models, thus leading to limited performances.

\subsection{Language Models for Recommendation}
As suggested in previous work~\cite{liu2023pre, wu2023survey, fan2023recommender,deldjoo2024review}, the adaption of large language models to the field of recommender systems can achieve strong abilities. Specifically, LLMs for recommendation can be classified as follows: LLM itself serves as a recommendation model~\cite{p5,m6rec,vats2024exploring} or LLM as a component for the recommendation model~\cite{kar,unisrec,huang2024foundation}.
LLMs can process complex natural language tasks, so one basic idea is to directly apply LLMs for recommendation tasks~\cite{tallrec,m6rec,chen2023large, zhu2023large}. For instance, P5~\cite{p5} utilizes the T5~\cite{t5} as a universal pre-trained model and unifies all recommendation tasks in the shared framework. M6-Rec~\cite{m6rec} extends the foundation model M6~\cite{m6} and supports open-ended domains and tasks in the industrial recommender system. TALLRec~\cite{tallrec} fine-tunes LLaMA-7B with the LoRA~\cite{lora} parameter efficient strategy and aligns it with recommendation. ReLLa~\cite{rella} further adopts retrieval-enhanced instruction tuning by adopting SUBR as a data augmentation technique for training samples and achieves better performance. 
But all the above methods suffer from inferring latency, so another direction is to combine LLMs and recommendation models~\cite{kar,xu2024prompting,li2023large}. For example, U-BERT~\cite{UBERT} utilizes BERT as a generator of user representations. UnisRec~\cite{unisrec} and VQRec~\cite{vqrec} feed BERT with descriptive texts and these output representations can be applied in cross domains. KAR~\cite{kar} adopts the GPT-3.5 to extract and reason knowledge from descriptive texts of users and items.

Although these models achieve higher performances by large language models, they suffer from two limitations: lifelong user behavior incomprehension problem, and the computational overhead due to the excessive executions of LLMs. Our proposed LIBER improves these methods by successfully solving such two limitations.

%% file: section/methodology.tex
\section{Methodology}
% We first provide an overview of our proposed Lifelong User Behavior Modeling Framework with Large Language Models (LLMs), and then elaborate on the details of each module. Finally, we discuss the efficiency of our proposed LIBER.
\subsection{Overview}
To utilize LLMs to assist the user behavior modeling, we design LIBER, as shown in Figure~\ref{fig: model}. The framework is model-agnostic and consists of the following three modules:

\textbf{User Behavior Streaming Partition Module} constructs a strategy to automatically divide the lifelong sequences of user behaviors into short-term behavior cache and long-term behavior memory, where the behaviors in the short-term behavior cache are only utilized by traditional recommendation method while the behaviors in the long-term behavior memory are employed by LLMs to assist the model. Besides, long-term behavior memory is conducted in fixed partitions, with new partitions being increased gradually only if certain partition conditions are met. The user behavior length in each partition is limited to an appropriate length for LLMs. In this way, both the lifelong user behavior incomprehension problem and the efficiency problem are well mitigated.

\textbf{User Interest Learning Module} extracts recommendation-relevant knowledge from LLMs for user behavior sequences. Each partition is processed to generate an intra-block user interest summary and an inter-block user interest shift in a cascading way. In this way, we can not only obtain the external knowledge to help the model understand the behaviors better in user sequences but also leverage powerful reasoning capability from LLMs to analyze how user interests evolve.

\textbf{User Interest Fusion Module} converts textual knowledge from LLMs into compact representations and fuses them properly to help the CTR prediction task. First, the knowledge obtained from LLMs is encoded into dense representations. Then, we design an attention fuse layer to integrate the user interests from different partitions. Finally, this long-term user interest fusion representation can be incorporated into the prediction by any recommendation model to improve the performance.

\subsection{User Behavior Streaming Partition}
Since the user behavior sequences are not only extremely lengthy but also frequently dynamically varied, as reported in section~\ref{sec:intro}, directly using LLMs to model the user behavior sequences suffers two serious problems: (1) Lifelong user behavior incomprehension problems; (2) The huge computational overhead due to the excessive executions of LLMs. 
Therefore, inspired by the computer data storage technique~\cite{cachememory}, we divide the lifelong user behavior sequences into short-term behavior cache and long-term behavior memory and only perform LLM enhancement for long-term behavior memory.

\renewcommand{\algorithmicrequire}{ \textbf{Input:}} 
\renewcommand{\algorithmicensure}{ \textbf{Output:}}

\begin{algorithm}[h]
    \caption{Incremental Behavior Partition Algorithm}
    \label{algo:training}
    \begin{algorithmic}[1]
        \REQUIRE
        User behavior streaming, short-term behavior cache, long-term behavior memory
        \ENSURE
        LLM-enhanced representation for long-term behavior memory
        \vspace{1mm}
        \STATE $j\leftarrow$ partition number of long-term behavior memory 
        % \FOR{$i\leftarrow 1$ to $t$}
        % \WHILE{True}
        \IF{a new behavior $b_i$ is received}
        \STATE Put $b_i$ into the short-term behavior cache
        \IF{partition condition is satisfied}
        \STATE $j \leftarrow j + 1$
        \STATE Add a new partition $\boldsymbol{P_j}$ in long-term behavior memory by the behaviors in cache
         \STATE empty the short-term behavior cache
        \STATE Generate user interest knowledge using LLMs
        \STATE Encode the knowledge as $\boldsymbol{r_j}$ according to Eq.~(\ref{eq:encode})
        \ENDIF
        \ENDIF
        % \ENDWHILE
        % \ENDFOR
        % \RETURN{ $\boldsymbol{r_1},\cdots,\boldsymbol{r_j}$}
    \end{algorithmic}
\end{algorithm}

\begin{figure*}[t]
  \centering
  \vspace{-0.8em}
  \includegraphics[width=\linewidth]{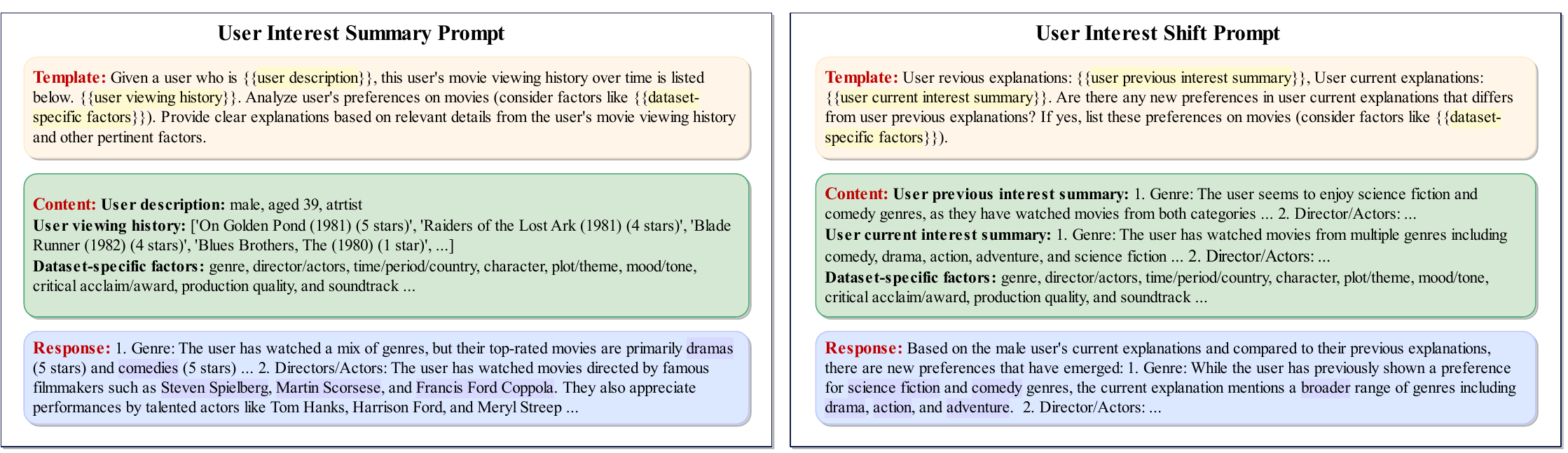}
  \vspace{-1.8em}
  \caption{Example prompts for LIBER. The yellow, green, and blue text bubbles represent the prompt template, the content to be filled in the template, and the response generated by LLMs respectively (some text has been omitted for page limitations).}
  \vspace{-1.0em}
  \label{fig: prompt_example}
\end{figure*}

Specifically, we propose an incremental behavior partition algorithm in Algorithm~\ref{algo:training}. Whenever a new user behavior $b_i$ is received, it is first allocated to the short-term behavior cache. Then we set a partition condition for this short-term behavior cache, which can be either time-related, number of behaviors related, or based on the more complicated automated model-based rule. In practice, we set a simple partition condition for convenience as 
\begin{equation}
    \operatorname{Len}\left(\textit{short-term behavior cahce}\right) \geq K~,
    \label{eq:condition}
\end{equation}
where $\operatorname{Len}\left(\cdot\right)$ denotes the number of behaviors and $K$ is a balance coefficient to control the partition condition strength.

When the partition condition is satisfied, all user behaviors in the cache are removed and utilized to construct a new partition $\boldsymbol{P_j}$ in long-term behavior memory. 
Then we make use of LLMs to extract additional external knowledge to get an LLM-enhanced representation for this partition, which will be illustrated in the next subsection.
In this way, since the past partitions for the same user are fixed and not changed in our method, we only need to execute LLMs for the new partition in memory, thus solving the computational overhead problem well. Additionally, the user sequence input for LLMs can be limited to a suitable length to avoid the lifelong user behavior incomprehension problem by our approach.

\subsection{User Interest Learning}
% In this subsection, we introduce our user interest learning module to both learn the user interest summary and user interest shift.

\subsubsection{User Interest Summary} LLMs have accumulated a great amount of factual knowledge from the web and possess strong reasoning abilities to analyze users' interests. Therefore, we can first employ LLMs to summarize and reason for user sequences. Different from traditional recommendation models that only memorize ID-based features through collaborative signals, LLMs can infer user interests from various perspectives through external semantic information based on user sequences. 

We design an instruction template $prompt_{s}$ for user interest summary (shown in Figure~\ref{fig: prompt_example}). $prompt_{s}$ is constructed with the user's profile description, behavior history and dataset-specific factors (\textit{i.e.,} genre, director, period for movie dataset), where the user profile description and behavior history provide LLMs with the necessary context and user-specific information to understand the user’s preferences while dataset-specific factors can instruct LLMs to analyze user preference from proper perspectives according to the specific dataset and allow LLMs to recall the relevant knowledge more comprehensively. 

Specifically, when a new partition $\boldsymbol{P_j}$ is generated, we utilize LLMs to yield user interest summary knowledge $klg_{j}^{s}$ as:
\begin{equation}
    klg_{j}^{s} = \text{LLM}\left(prompt_{s}, \boldsymbol{P_j}\right)~.
    \label{eq:summary}
\end{equation}
% where $prompt_{s}$ is the instruction template prompt for user interest summary.

\subsubsection{User Interest Shift}
Although the difficulty for LLMs to understand the user sequences is reduced when we employ the partition strategy, it still exists another serious challenge: hard to understanding the dynamic variations of user interests for LLMs, which is even exacerbated by the fact that partitioning cuts off the connection within the user sequences. 
To solve this problem, we propose a cascaded paradigm, trying to learn how user interests evolve.

Here we design a user interest shift instruction template $prompt_{c}$. As shown in Figure~\ref{fig: prompt_example}, $prompt_c$ consists of three parts - user previous interest summary, user current interest summary, and dataset-specific factors. This prompt can guide LLMs to pay more attention to users' new interests and eliminate the obsolescent interests, to better explore and understand the changes in user interests.

Specifically, for the new partition $\boldsymbol{P_j}$, after the user interest summary knowledge $klg_{j}^{s}$ is got, we can employ LLMs to get user interest shift knowledge $klg_{j}^{c}$ as:

\begin{equation}
    klg_{j}^{c} = LLM(prompt_c, klg_{j}^{s}, klg_{j-1}^{s})~.
    \label{eq:shift}
\end{equation}
% where $prompt_{c}$ is the instruction template prompt for user interest shift and $klg_{j-1}^{s}$ is the user interest summary knowledge for the partition $\boldsymbol{P_{j-1}}$.
% where $klg_{j-1}^{s}$ is the user interest summary knowledge for the partition $\boldsymbol{P_{j-1}}$.

\subsection{User Interest Fusion}
The knowledge generated by LLMs brings new challenges when used to assist recommendation models: (1) The knowledge generated by LLMs is usually in textual form, which cannot be directly utilized by traditional recommender systems. (2) Our partition approach yields multiple partitions of knowledge, and how to combine the knowledge from different partitions becomes a new challenge.

To address these challenges, we devise the user interest fusion module with three components: a \textit{knowledge encoder}, a \textit{ attention fuse layer}, and a \textit{combination layer}. 
% First, the knowledge encoder encodes the generated textual knowledge for each partition into dense representations. Then, the attention fuse layer merges the dense representations from different partitions into a unified fusion representation. Finally, the combination layer incorporates this fusion representation into the traditional recommender system.

\subsubsection{Knowledge Encoder}
To utilize the semantic information generated by LLMs, we propose a knowledge encoder to encode these texts outputted by LLMs. The knowledge encoder is a smaller language model that has fewer parameters than LLMs, so it is suitable to vectorize the text quickly. 
We can obtain the dense representation $\boldsymbol{r_j}$ for partition $\boldsymbol{P_j}$ as:
\begin{equation}
    \boldsymbol{r_j} = \operatorname{Encoder}(klg_j^s,klg_j^c)~,
    \label{eq:encode}
\end{equation}
where $klg_j^s$ and $klg_j^c$ denote the user interest summary knowledge and user interest shift knowledge generated by LLMs for $j$-th partition. $\operatorname{Encoder}$ is a low-parameterized language model, in practice, we primarily use BERT~\cite{bert}. Note that the output dimension for BERT is too large (768 for each token), which does not match the traditional recommender system, here we additionally apply a pre-trained PCA~\cite{pca} to do the dimensionality reduction.

\subsubsection{Attention Fuse Layer} 
To fuse the representations from different partitions to assist the recommendation, we utilize the self-attention pooling~\cite{attention} to fuse these representations.  
Suppose the user long-term behavior memory contains $j$ partitions $\left\{\boldsymbol{P_1},\cdots, \boldsymbol{P_j}\right\}$, the fusion representations $\hat{\boldsymbol{r_j}}$ can be computed as:
\begin{equation}
    \hat{\boldsymbol{r_j}} = \operatorname{SelfAttn}(\boldsymbol{r_1},\dots,\boldsymbol{r_j})=\operatorname{SelfAttn}(\mathbf{R_j})~,
\end{equation}
where $\mathbf{R_j}$ is the concatenation of $\left\{\boldsymbol{r_1},\dots,\boldsymbol{r_j}\right\}$ and the $\operatorname{SelfAttn}$ calculates the weights of different partitions and obtains a unified weighted representation.
The detailed process can be defined as:
\begin{equation}
    \operatorname{SelfAttn}(\mathbf{R}) = \operatorname{Aggr}\left(\operatorname{Softmax}\left(\frac{\mathbf{R}\mathbf{W}^Q(\mathbf{R}\mathbf{W}^K)^T}{\sqrt{d}}\right)\mathbf{R}\mathbf{W}^V\right)~,
\end{equation}
where $\mathbf{W}^{Q},\mathbf{W}^{K},\mathbf{W}^{V}$ denote the weights of query, key and $d$ is the output dimension of $ \operatorname{SelfAttn}(\mathbf{R})$.

\subsubsection{Combination Layer}
Once we obtain the LLM-enhanced user interest representation, we can incorporate it into the traditional recommendation model. Specifically, we put it as an additional feature in the recommendation. Generally, it can be formulated as:
\begin{equation}
    \hat{y} = f(x, h, \hat{\boldsymbol{r}}; \theta)~,
\end{equation}
where $x$ is the general user and item feature, $h$ is the user history feature and $\hat{\boldsymbol{r}}$ is the long-term user interest representation. Importantly, although $\hat{\boldsymbol{r}}$ may ignore some short-term behaviors since they have not been added to the long-term behavior memory and are still present in the short-term behavior cache, these behaviors are modeled in the traditional way using $h$ to capture the short-term user interest, resulting in a better balance between effectiveness and efficiency. Besides, LIBER only modifies the input of the recommendation model, thus it is a model-agnostic framework with various backbone recommendation model designs.

\subsection{Discussion of Efficiency}
Compared with the methods directly using LLMs to enhance the user sequence modeling~\cite{kar,rella}, LIBER can reduce the time cost of executing LLMs, which accounts for the majority of offline time. Suppose the sample number in the dataset is $N$ and the time complexity of executing LLMs once is $O(M)$, the other LLM-enhanced works for user sequence modeling need to employ LLMs once for each sample and therefore need cost $O(N\times M)$ for LLMs execution. However, for LIBER, we design the short-term behavior cache and long-term behavior memory in an incremental paradigm. Therefore, different samples from the same user can share the output of LLMs. Suppose the average length for partition is $K$, then the time complexity of LIBER to executing LLMs is $O(N\times M / B)$, which mitigates the computational overhead problem greatly. 

For latency requirements in online scenarios, LIBER can pre-store the representations in a database. Consequently, we only need to retrieve the representations from the database for the inference process, which makes the inference time of LIBER acceptable.

%% file: section/experiment.tex
\begin{table*}
\vspace{-0.5em}
\caption{The overall performance on MovieLens-100K and Amazon-books datasets. The best result is given in bold, and the second-best value is underlined. "Rel.Impr." is the relative improvement rate of LIBER against each baseline. }
\vspace{-0.5em}
\label{table:all}
\begin{threeparttable}[t]{
\begin{tabular}{cccccc|cccc}
\toprule
{\multirow{2}{*}{\textbf{Model}}} & {\multirow{2}{*}{\textbf{Language Model}}} & \multicolumn{4}{c|}{\textbf{MovieLens-100K}} & \multicolumn{4}{c}{\textbf{Amazon-books}} \\ 
\cmidrule(lr){3-6}\cmidrule(lr){7-10}
 &{} & {\textbf{AUC}}& {\textbf{Rel.Impr.}} &{\textbf{Log Loss}}& {\textbf{Rel.Impr.}} & {\textbf{AUC}}& {\textbf{Rel.Impr.}}  &{\textbf{Log Loss}}& {\textbf{Rel.Impr.}}\\ 
\midrule
{DCN}&{-} & {0.7932}&{3.69\%}&{0.5288}&{7.19\%}& {0.8341}&{1.59\%}&{0.4968}&{4.51\%} \\ 
{DeepFM}&{-}  & {0.7919}&{3.86\%}&{0.5296}&{7.33\%}& {0.8335}&{1.67\%}&{0.4982}&{4.78\%} \\
{xDeepFM}&{-} & {0.7937}&{3.63\%}&{0.5285}&{7.13\%}& {0.8346}&{1.53\%}&{0.4965}&{4.45\%} \\
{AutoInt}&{-} & {0.7944}&{3.54\%}&{0.5282}&{7.08\%}& {0.8348}&{1.51\%}&{0.4969}&{4.53\%} \\
{FiBiNet}&{-} & {0.7899}&{4.13\%}&{0.5308}&{7.54\%}& {0.8318}&{1.88\%} &{0.4989}&{5.00\%}  \\
{DIN} &{-}& {0.7956}&{3.38\%}&{0.5259}&{6.67\%}& {0.8357}&{1.40\%}&{0.4954}&{4.24\%} \\
{DIEN} &{-}& {0.7988}&{2.97\%}&{0.5178}&{5.21\%}&{0.8367}&{1.28\%}&{0.4932}&{3.81\%}\\
\midrule
{UnisRec}&{BERT-100M} &{0.7921}&{3.84\%}&{0.5294}&{7.29\%}&{0.8217}&{3.13\%}&{0.5025}&{5.59\%}\\
{VQRec}&{BERT-100M} & {0.7929}&{3.73\%}&{0.5290}&{7.22\%}&{0.8251}&{2.70\%}&{0.5099}&{6.69\%}\\
{PTab} &{BERT-100M}& {0.7962}&{3.30\%}&{0.5214}&{5.87\%}&{0.8359}&{1.38\%}&{0.4953}&{4.22\%}\\
{P5}&{T5-223M} & {0.7958}&{3.36\%}&{0.5253}&{6.57\%}&{0.8382}&{1.10\%}&{0.4911}&{3.40\%}\\
{TALLRec}&{LLaMa2-13B} & \underline{0.8076}&{1.84\%}&\underline{0.5088}&{3.54\%}&{0.8409}&{0.77\%}&{0.4802}&{1.21\%}\\
{KAR}&{LLaMa2-13B} & {0.8072}&{1.90\%}&{0.5132}&{4.56\%}&\underline{0.8422}&{0.62\%}&\underline{0.4794}&{1.04\%}\\
\midrule
{LIBER}&{LLaMa2-13B} & {\textbf{0.8225$^*$}}&{-}&{\textbf{0.4908$^*$}}&{-}&{\textbf{0.8474$^*$}}&{-}&{\textbf{0.4744$^*$}}&{-}\\   
   \bottomrule          
\end{tabular}
}
\end{threeparttable}
 \begin{tablenotes}
		 \item   \small $^*$ denotes statistically significant improvement (measured by t-test with p-value$<$0.001) over baselines.
	\end{tablenotes}
 \vspace{-0.3em}
\end{table*}

\section{EXPERIMENTS}\label{sec:exp}
To gain more insights into LIBER, we tend to address the following research questions (RQs) in this section.

 \begin{itemize}
[leftmargin = 12 pt]
 \item \textbf{RQ1}: How does LIBER perform compared with current LLM-enhanced and traditional recommendation models?
 \item \textbf{RQ2}: Does LIBER gain performance improvements for different backbone recommendation models and large language models?
 \item \textbf{RQ3}: Can LIBER improve the performance of existing models in a live recommender system?
 \item \textbf{RQ4}: What are the influences of different components in LIBER?
 % \item \textbf{RQ5}: How about the time complexity of LIBER compared with other models?\
  \item \textbf{RQ5}: How about the time complexity of LIBER?
 \end{itemize}
\subsection{Experimental Settings}

\subsubsection{Dataset} 
Experiments are conducted for the following two public datasets and one industrial dataset:

 \begin{itemize}[leftmargin = 10 pt]
\item \textbf{MovieLens-100K.}\footnote{\url{https://grouplens.org/datasets/movielens/100k/}} This is a movie review dataset, which contains 100,000 samples with 910 users and 3310 items. According to the setting of sequential CTR prediction~\cite{din,tallrec}, we convert the rating score above 3 as the positive, and the rest as the negative. The dataset is split into training and testing sets with a ratio of 9:1 according to the global timestamp. 
\item \textbf{Amazon-Books.}\footnote{\url{https://cseweb.ucsd.edu/~jmcauley/datasets/amazon_v2/}} This is the Amazon Books Review Dataset. Following the data processing similar to~\cite{kar}, we filter out the less-interacted users and items, retaining 11,906 users and 17,332 items with 1,406,582 samples. The ratings of 5 are regarded as positive and the rest as negative. The other preprocessing is similar to MovieLens-100K.
 \end{itemize}

\subsubsection{Baselines and Evaluation Metrics}
To evaluate the superiority and effectiveness of our proposed model, we compare LIBER with two classes of existing models: (1) \textit{traditional CTR models} that take one-hot encoded ID features as inputs, which can be further divided into feature interaction models (DCN~\cite{dcn}, DeepFM~\cite{deepfm}, xDeepFM~\cite{xdeepfm}, AutoInt~\cite{autoint}, FiBiNet~\cite{fibinet}) and user behavior models (DIN~\cite{din}, DIEN~\cite{dien}); and (2) \textit{LM-based models} that leverage language model to enhance recommendation, containing UnisRec~\cite{unisrec}, VQRec~\cite{vqrec}, PTab~\cite{ptab}, P5~\cite{p5}, TALLRec~\cite{tallrec} and KAR~\cite{kar}.

The common evaluation metrics for CTR prediction are \textbf{AUC} (area under the ROC curve) and \textbf{Log Loss} (cross-entropy).

% \subsubsection{Evaluation Metrics} To evaluate the performance of our method, we utilize AUC (area under the ROC curve) and Log Loss (binary cross-entropy loss) as evaluation metrics following~\cite{din,deepfm,autoint}. A higher AUC or a lower Log Loss, even by a small margin (\textit{e.g.}, 0.001) can be viewed as a significant improvement~\cite{xdeepfm,dcnv2}.

\begin{table*}
% \vspace{-0.5em}
\caption{Compatibility analysis of LIBER for different backbone CTR models on MovieLens-100K and Amazon-books datasets. "Rel.Impr." is the relative improvement rate of LIBER against each baseline.}
\vspace{-0.5em}
\resizebox{\textwidth}{!}{
\begin{tabular}{ccccccc|cccccc}
\toprule
{\multirow{3}{*}{\makecell{\textbf{Backbone}\\ \textbf{Model}}}} & \multicolumn{6}{c|}{\textbf{MovieLens-100K}} & 
\multicolumn{6}{c}{\textbf{Amazon-books}} \\ 
\cmidrule{2-13}
\multicolumn{1}{c}{} & \multicolumn{3}{c}{\textbf{AUC}} & \multicolumn{3}{c|}{\textbf{Log Loss}} & \multicolumn{3}{c}{\textbf{AUC}} & \multicolumn{3}{c}{\textbf{Log Loss}}\\ 
\cmidrule{2-13}
\multicolumn{1}{c}{} & \textbf{Base} & \textbf{LIBER} & \textbf{Rel.Impr.} & \textbf{Base} & \textbf{LIBER} & \textbf{Rel.Impr.} & \textbf{Base} & \textbf{LIBER} & \textbf{Rel.Impr.} & \textbf{Base} & \textbf{LIBER} & \textbf{Rel.Impr.} \\ 
\cmidrule{1-13}
{DCN} & 0.7932  & \textbf{0.8112$^*$}  &  2.27\% & 0.5288 & \textbf{0.5091$^*$} & 3.72\% & 0.8341  & \textbf{0.8418$^*$}  &  0.92\% & 0.4968 & \textbf{0.4795$^*$} & 3.61\% \\ 
{DeepFM} & 0.7919  & \textbf{0.8137$^*$}  &  2.75\% & 0.5296 & \textbf{0.5073$^*$} & 4.21\% & 0.8335  & \textbf{0.8395$^*$}  &  0.72\% & 0.4982 & \textbf{0.4887$^*$} & 1.94\% \\ 
{xDeepFM} & 0.7937  & \textbf{0.8148$^*$}  &  2.66\% & 0.5285 & \textbf{0.5047$^*$} & 4.50\% & 0.8346  & \textbf{0.8412$^*$}  &  0.79\% & 0.4965 & \textbf{0.4803$^*$} & 3.37\% \\ 
{AutoInt} & 0.7944  & \textbf{0.8150$^*$}  &  2.59\% & 0.5282 & \textbf{0.5042$^*$} & 4.54\% & 0.8348  & \textbf{0.8423$^*$}  &  0.90\% & 0.4969 & \textbf{0.4784$^*$} & 3.87\% \\ 
{FiBiNet} & 0.7899  & \textbf{0.8107$^*$}  &  2.63\% & 0.5308 & \textbf{0.5098$^*$} & 3.96\% & 0.8318  & \textbf{0.8385$^*$}  &  0.81\% & 0.4989 & \textbf{0.4911$^*$} & 1.59\% \\ 
{DIN} & 0.7956  & \textbf{0.8154$^*$}  &  2.49\% & 0.5259 & \textbf{0.5037$^*$} & 4.22\% & 0.8357  & \textbf{0.8433$^*$}  &  0.91\% & 0.4954 & \textbf{0.4767$^*$} & 3.92 \% \\ 
{DIEN} & 0.7988  & \textbf{0.8225$^*$}  &  2.97\% & 0.5178 & \textbf{0.4908$^*$} & 5.21\% & 0.8367  & \textbf{0.8474$^*$}  &  1.28\% & 0.4932 & \textbf{0.4744$^*$} & 3.96\% \\  
   \bottomrule          
\end{tabular}
}
 \begin{tablenotes}
 \item  \small $^*$ denotes statistically significant improvement (measured by t-test with p-value$<$0.001) over baselines.
\end{tablenotes}
\vspace{-0.5em}
\label{table:compatibility1}
\end{table*}

 \subsubsection{Implementation Details}
We select the LLaMa2-13B~\cite{llama2} released by Meta for LIBER. Then, BERT~\cite{bert} is employed to encode the knowledge, followed by average pooling as the aggregation function. The partition condition balance coefficient in Eq.~(\ref{eq:condition}) is set as 20. Since our framework is model-agnostic, we utilize the best traditional recommendation baseline DIEN as our backbone model. We keep the embedding size as 32, and the output layer MLP size as $[200,80]$. The batch size is searched from $\{512, 1024\}$ and learning rate is tuned from $\{1e-3, 7.5e-4, 5e-4, 2.5e-4, 1e-4\}$. For fair comparisons, the parameters of all the baselines are also tuned by grid search to achieve optimal performances.
Besides, the behavior sequence length utilized by LLMs does not exceed the behavior sequence length used by the baselines for fairness.

% Note for the industrial dataset, we do not compare the user behavior model (DIN), because we have added the attention mechanism on sequences to each baseline model, which is the core of DIN. The backbone model for LIBER on the industrial dataset is FiBiNet.

\subsection{Performance Comparison (RQ1)}
In this section, we compare the performance of LIBER with the baseline models. Table~\ref{table:all} summarizes the performance on MovieLens-100K and Amazon-books datasets. We can observe:

\begin{itemize}[leftmargin = 10 pt]
    \item \textbf{User behaviors sequence modeling can significantly improve performance.} For instance, DIN and DIEN, achieve better performances than other traditional models in terms of AUC and Log Loss, which validates the importance of modeling user behavior sequences for CTR prediction.
    \item \textbf{Leveraging large language models (LLMs) brings benefit to model performance.} TALLRec and KAR, utilizing LLaMa2-13B, achieve the best performances among all the baseline models. However, the models using language models (LMs) with parameters less than one billion, such as UnisRec, VQRec, PTab, and P5, tend to exhibit poorer performances, even worse than the traditional CTR models. These results demonstrate the large ability gap between the LMs and LLMs. 
    \item \textbf{The superior performance of LIBER.} We can observe from Table~\ref{table:all} that LIBER consistently yields the best performance on all datasets. Concretely, LIBER beats the best baseline by \textbf{1.90\%} and \textbf{4.56\%} on MovieLens-100K dataset in terms of AUC and Log Loss (\textbf{0.62\%} and \textbf{1.04\%} on Amazon-books dataset). This suggests that modeling lifelong user behavior sequences by large language models can greatly improve CTR prediction performance.
\end{itemize}

\subsection{Compatibility Analysis (RQ2)}
% In this section, we investigate the compatibility of LIBER with different backbone recommendation models and different large language models, respectively.

\subsubsection{The Compatibility for Backbone Recommendation Model}
To investigate LIBER's compatibility for different backbone recommendation models, we implement our proposed LIBER upon representative CTR models for MovieLens-100K and Amazon-books datasets. The results are presented in Table~\ref{table:compatibility1}, from which the following observations can be made:

\begin{itemize}[leftmargin = 10 pt]
    \item As a model-agnostic framework, LIBER can achieve performance improvements for different backbone recommendation models. With the help of LIBER, the selected representative CTR models on two datasets both achieve a significant AUC improvement, indicating the compatibility of LIBER.
    \item For feature interaction models and user behavior models, the relative improvements by LIBER are similar. This may be because LIBER mostly considers the semantic information whereas the user behavior models account for the collaborative information, so there is only little information overlap between LIBER and user behavior models. Therefore, the enhancements to feature interaction models and user behavior models are similar.
\end{itemize}

\begin{table}[t]
% \vspace{-0.3em}
\caption{Compatibility analysis of LIBER for different Large Language Models on MovieLens-100K Dataset. "Rel.Impr." is the relative improvement rate of LIBER against DIEN.}
\vspace{-0.7em}
\centering
\resizebox{\linewidth}{!}{
\begin{tabular}{ccccc}
\toprule
\textbf{Model} & \textbf{AUC} & \textbf{Rel.Impr.}  & \textbf{Log Loss} & \textbf{Rel.Impr.}\\ 
\midrule
{DIEN}  & 0.7988 & - &  0.5178  & -\\
LIBER (ChatGLM2-6B) & 0.8142 & 1.93\% & 0.5044 & 2.59\% \\
LIBER (Llama2-7B) & 0.8177 & 2.37\% & 0.4981 & 3.80\% \\
LIBER (Llama2-13B) & 0.8225 & 2.97\% & 0.4908 & 5.21\% \\
   \bottomrule          
\end{tabular}
}
\vspace{-1.0em}
\label{table: llm}
\end{table}

\subsubsection{The Compatibility for Large Language Models}
To demonstrate the compatibility of LIBER for different large language models, we implement our proposed LIBER upon three representative LLMs~\cite{glm,llama,llama2} on the MovieLens-100K dataset. As shown in Table~\ref{table: llm}, we can get several conclusions:
\begin{itemize}[leftmargin = 10 pt]
    \item By all three large language models, employing LIBER can improve the performance, which validates that LIBER has a strong compatibility for different LLMs.
    \item The large language models with more parameters enable LIBER to achieve better performance. It may be attributed to the larger LLMs having better comprehension and reasoning capabilities, beneficial for understanding the user interest from the lifelong behavior sequences.
\end{itemize}

\subsection{Industrial Experiments (RQ3)}
\subsubsection{Offline Industrial Experiments} We also test our approach on an offline industrial dataset collected and sampled from a music recommendation scenario, where hundreds of millions of samples are generated. We split them into training/test sets with a ratio of $9:1$ by timestamp. We utilize around $100$ feature fields, including user features (\textit{e.g.}, user's behavior history), music features (\textit{e.g.}, artist, category), and context features.
Note for this industrial dataset, we do not compare the user behavior model (DIN), because we have applied the attention mechanism on sequential features to each baseline model, which is the core of DIN.

 \begin{table}[h]
 % \vspace{-0.2em}
\caption{The performance over Industrial Dataset. Note that the attention mechanism on user sequence (DIN) has been applied to each model.}
\vspace{-0.5em}
\centering
\begin{tabular}{ccccc}
\toprule
{ \textbf{Model}}& {\textbf{AUC}}  & {\textbf{Rel.Impr.}} & {\textbf{Log Loss}} & {\textbf{Rel.Impr.}} \\ 
\midrule
{DCN} & 0.7351 & 0.76\% & 0.3328  & 0.30\% \\
{DeepFM} & 0.7336 & 0.97\% & 0.3335 & 0.51\% \\
{AutoInt} & 0.7334 & 1.00\% & 0.3336 & 0.54\% \\
{FiBiNet} &  0.7362 & 0.61\% & 0.3325 & 0.21\% \\
{KAR} & \underline{0.7380} & 0.37\% & \underline{0.3323}  &   0.15\% \\
{LIBER} & \textbf{0.7407$^*$} & - & \textbf{0.3318$^*$} & - \\
   \bottomrule          
\end{tabular}
\label{table: ads}
 \begin{tablenotes}
 \item  \small  $^*$ denotes statistically significant improvement (p-value$<$0.001).
\end{tablenotes}
\vspace{-0.5em}
\end{table}

From Table~\ref{table: ads}, we summarize the observations in two points. Firstly, the methods employing large language models can consistently improve the performance, which shows the importance of introducing language models. Secondly, our proposed LIBER outperforms all the other models, demonstrating the value of modeling lifelong user behavior sequences by large language models.

\subsubsection{Online Industrial A/B Test}
To validate the effectiveness of our approach in the real environment, we deploy LIBER on Huawei's music recommendation service and conduct a one-week online A/B test. Specifically, 10\% of users are randomly selected into the experimental group, and another 10\% are in the control group. For the control group, the users are served by a highly optimized deep model. For the experimental group, the users are served by the same base model with LIBER, utilizing Huawei’s own LLM Pangu~\cite{pangu} to generate lifelong user preference. It is worth noticing that LIBER is updated in a streaming paradigm, where the new user behavior is firstly saved to a short-term behavior cache. When the size of this cache exceeds 20, we take out these behaviors to construct a new partition in long-term behavior memory and then make use of LLM to extract user interest summary and shift. In this way, the number of LLM calls is greatly reduced and can be acceptable for our online scenario. We compare the performances in terms of user play count and play time, which are also widely used in industrial music services products~\cite{kar,dai2024modeling}.

\begin{table}[h]
\vspace{-0.2em}
	\caption{LIBER's Lift rate of online A/B test.}
 \vspace{-0.5em}
	\label{tab:online}
	\centering
		\begin{tabular}{ccc}
			\toprule
            Metric & Play Count & Play Time \\ \midrule
            Lift rate & $+3.01\%$  & $+7.69\%$ \\
			\bottomrule
		\end{tabular}
  \vspace{-0.5em}
\end{table}

Table~\ref{tab:online} shows the improvement of LIBER over the base model, where LIBER achieves substantial improvements in user's play count and play time by 3.01\% and 7.69\%, respectively. It is a significant improvement and demonstrates the magnificent effectiveness of our proposed model.

\subsection{Ablation Study (RQ4)}
To investigate the effectiveness of each component in our proposed LIBER framework, we design the following model variants:
 \begin{itemize}[leftmargin=10pt]
    \item \textbf{LIBER (Ours)} is the complete version of our proposed method with three key components: User Behavior Streaming Partition, User Interest Learning, and User Interest Fusion.
    \item \textbf{LIBER (w/o Part.)} removes the user behavior streaming partition module and takes the original long historical sequence as inputs for LLMs to enhance the user preference modeling for backbone models.
    \item \textbf{LIBER (w/o I.S.)} removes the user interest shift part in the user interest learning module, which is designed to address the ever-evolving user dynamic interest via a cascaded merging paradigm among user behavior partitions.
    \item \textbf{LIBER (w/o A.F.)} replaces the attention fuse layer in the user interest fusion module with a simple mean pooling layer.
 \end{itemize}

 \begin{figure}[h]
     \centering
     \vspace{-0.2em}
    \includegraphics[width=1.0\linewidth]{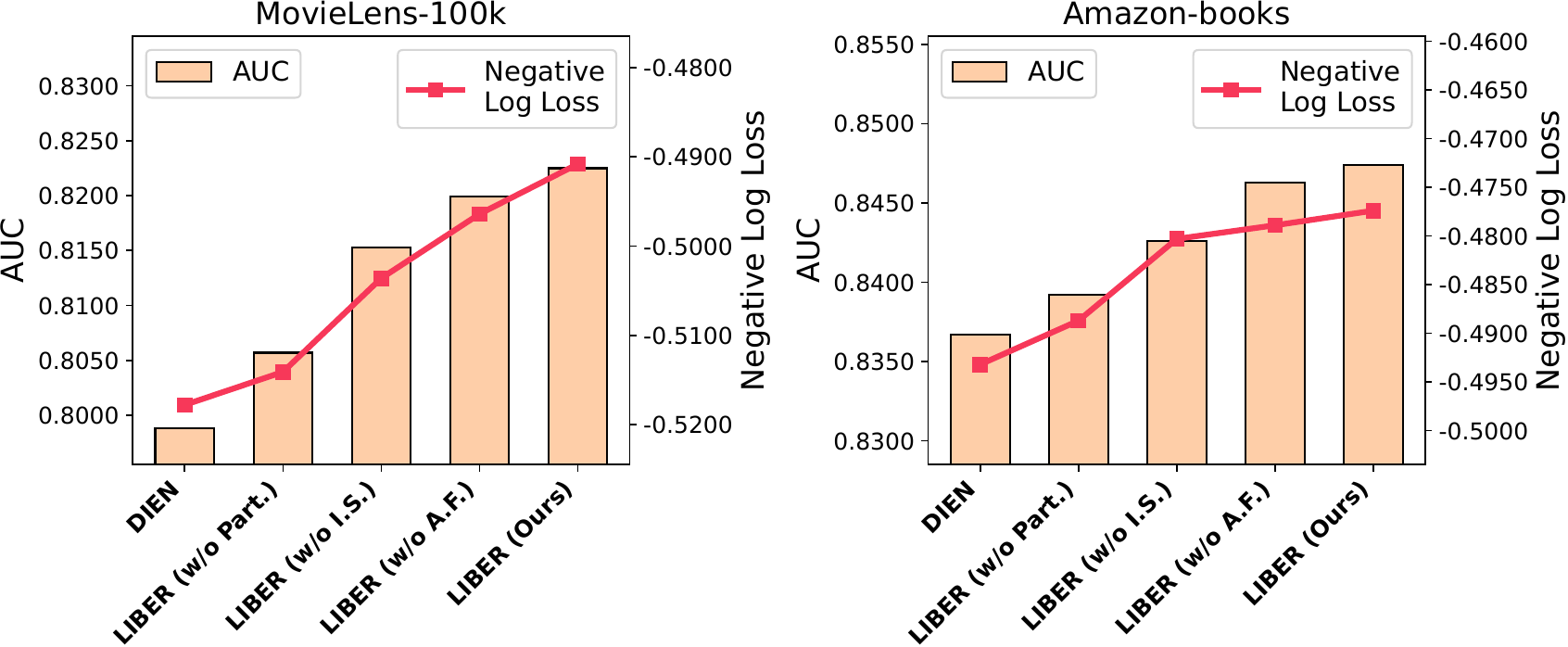}
    \vspace{-1.5em}
        \caption{Ablation study about the effectiveness of different components in LIBER.}
        \vspace{-0.9em}
        \label{fig:ablation}
\end{figure}

We adopt DIEN~\cite{dien} as the backbone model, and report the results in Figure~\ref{fig:ablation}, from which we draw the following observations:

\begin{itemize}[leftmargin=10pt]
    \item The performance of LIBER largely degenerates when the user behavior partition module is removed, \textit{i.e.,} LIBER (w/o Part.). By dividing the lifelong historical sequence into memory partitions, LIBER can help LLMs to well comprehend and extract useful semantic knowledge about user preferences, thereby yielding superior performance enhancements.
    \item The performance drops for LIBER (w/o I.S.), where we remove the user interest shift part in the user interest learning module. 
    This validates that it is essential to connect different memory partitions in a cascaded prompting way, which enables LLMs to capture the ever-evolving user dynamics for long-term preference modeling.
    \item The performance also degrades for LIBER (w/o A.F.). This suggests that it is necessary to employ the self-attention operator to aggregate the knowledge representations from different memory partitions, which contribute differently to the final user interest modeling.
\end{itemize}

\subsection{Efficiency Study (RQ5)}

In this section, we study the efficiency of LIBER in both the training phase and inference phase on the MovieLens-100K dataset, with DIEN~\cite{dien} as the backbone recommendation model.

\subsubsection{Training Efficiency}
For the training phase, the major bottleneck lies in the involvement of LLMs. Hence, in Table~\ref{tbl: traintime}, we report the \textbf{AUC} performance, the number of averaged LLM calls per user (\textbf{\#Calls/User}), the number of averaged tokens per prompt (\textbf{\#Tokens/Prompt}), and the averaged LLM processing time per user (\textbf{Time/User}) for the four model variants:
\begin{itemize}[leftmargin=10pt]
    \item \textbf{LIBER (Ours)} is the complete version of our proposed model, which involves both the user interest summary prompt and the user interest shift prompt.
    \item \textbf{LIBER (w/o I.S.)} removes the user interest shift prompt and only maintains the user interest summary prompt for each individual memory partition.
    \item \textbf{LLM (length=100)} further removes the partition-wise user interest summary prompt and employs LLMs to infer the user preference at each time step of updating the behavior sequence. We truncate the maximum length of the user sequence to 100, which is the same as LIBER.
    \item \textbf{LLM (length=20)} is similar to LLM (length=100), while we truncate the maximum sequence length to 20.
\end{itemize}

\begin{table}[t]
\vspace{-0.2em}
\caption{The training efficiency on MovieLens-100K dataset. We report the AUC performance, the number of averaged LLM calls per user (\#Calls/User), the number of averaged tokens per prompt (\#Tokens/Prompt), and the averaged LLM processing time per user (Time/User).}
\vspace{-0.7em}
\centering
\renewcommand{\arraystretch}{1.15}
\resizebox{1.0\linewidth}{!}{
\begin{tabular}{ccccc}
\toprule
\textbf{Variants}& \textbf{AUC}  & \textbf{\#Calls/User} & \textbf{\#Tokens/Prompt} &\textbf{Time/User (s)} \\ 
\midrule
{LIBER (Ours)} &{0.8225}&{8.96}&{1246.6}&{227.8}\\
{LIBER (w/o I.S.)} &{0.8153}&{4.98}&{967.1}&{115.0} \\
{LLM (length=100)} &{0.8057}&{89.06}&{1958.9}&{2223.8} \\
{LLM (length=20)} &{0.8083}&{89.06}&{967.1}&{2056.4} \\
   \bottomrule          
\end{tabular}
}
\vspace{-0.7em}
\label{tbl: traintime}
\end{table}

From Table~\ref{tbl: traintime}, we can obtain the following observations:
\begin{itemize}[leftmargin=10pt]
    \item Comparing with LLM (length=20), LLM (length=100) meets a significant performance degeneration, and meanwhile suffers from the severe training inefficiency for user interest modeling with LLMs. This validates the lifelong user behavior incomprehension problem~\cite{rella} and yields the necessity of our proposed LIBER framework for efficient lifelong sequence modeling.
    \item LIBER achieves better performance compared to vanilla LLM processing, with a relatively lower training cost. This verifies the superiority of LIBER in not only improving the user modeling efficiency with partition and cascaded prompting but also enhancing the recommendation performance by promoting lifelong sequence comprehension.
    \item Comparing LIBER with LIBER (w/o Interest Shift), although the introduction of the user interest shift prompt would double the training cost of LIBER with the cascaded merging paradigm, it is essential to effectively capture the ever-evolving user dynamics and improve the user preference estimation.
\end{itemize}

\begin{table}[h]
% \vspace{-0.7em}
\caption{The inference time per 1000 samples of different models on MovieLens-100K dataset.}
\vspace{-0.7em}
\centering
\renewcommand{\arraystretch}{1.15}
\resizebox{0.4\textwidth}{!}{
\begin{tabular}{ccccc}
\toprule
{\textbf{Model}} & {TALLRec}&{DIEN}&{KAR}&{LIBER}\\ 
\midrule
\textbf{Time (s)} &{9.96}&{$8.38 \times 10^{-2}$}&{$9.12 \times 10^{-2}$} &{$9.61 \times 10^{-2}$} \\
   \bottomrule          
\end{tabular}
}
\vspace{-0.7em}
\label{tbl: infertime}
\end{table}

\subsubsection{Inference Efficiency}
For the inference phase, we report the inference time per 1000 samples of TALLRec~\cite{tallrec}, DIEN~\cite{dien}, KAR~\cite{kar}, and our proposed LIBER in Table~\ref{tbl: infertime}. 
We can observe that TALLRec generally suffers from an unacceptable inference latency since it directly employs LLMs to estimate the user preference toward each target item.
Similar to KAR~\cite{kar}, LIBER can pre-calculate and cache the knowledge representations for long-term user interests, and therefore avoid LLM calls during the inference phase. 
Consequently, LIBER is able to achieve superior performance over baselines like DIN and KAR, and meanwhile satisfy the strict latency constraint in real-world applications.

%% file: section/conclusion.tex
\section{conclusion}
In this work, we propose \textbf{Li}felong User \textbf{Be}havio\textbf{r} Modeling (LIBER) based on large language models (LLMs). LIBER aims to solve two main challenges: (1) Lifelong user behavior incomprehension problem; and (2) Huge computational overhead of LLMs. LIBER mainly consists of three stages: User Behavior Streaming Partition (UBSP), User Interest Learning (UIL), and User Interest Fusion (UIF). UBSP condenses lengthy user behavior sequences into shorter partitions in an incremental paradigm. UIL leverages LLMs in a cascading way to infer insights from each partition. UIF integrates the textual outputs generated by LLMs to construct a unified representation and then incorporates this representation into the recommendation models. 
The significant improvements in offline evaluations and an online A/B test have demonstrated the superiority of our model.

% To validate the effectiveness of our proposed LIBER, we conducted rigorous experiments utilizing three datasets (two public and one industrial), demonstrating its superior performance compared to existing state-of-the-art baselines.

%% file: section/appendix.tex
% \begin{figure*}[h]
%   \centering
%   \includegraphics[width=\linewidth]{figure/prompt_v2.pdf}
%   \caption{Example Prompt of LIBER.}
%   \label{fig: prompt_example}
% \end{figure*}